\documentclass[twocolumn,superscriptaddress,nofootinbib, floatfix, amsfonts, aps]{revtex4} 
\usepackage{amsmath}
\usepackage{bm}
\usepackage{graphicx}
\usepackage{subfigure}
\usepackage{float}
\usepackage{mathrsfs}
\usepackage{multirow}
\usepackage{graphicx}
\usepackage{booktabs, ctable}
\usepackage{soul,xcolor}
\usepackage{xcolor, ulem, color, framed}
\usepackage{amssymb}

\newcommand{\sred}{\textcolor{black}}

\usepackage[colorlinks, linkcolor=red,anchorcolor=green,citecolor=blue]{hyperref}

\graphicspath{{./figs/}}

\begin{document}

 \title{Charmonium transport in the high-$\mu_B$ medium}
 
 \author{Jiaxing Zhao}
\affiliation{Department of Physics, Tsinghua University, Beijing 100084, China}
\affiliation{SUBATECH, Universit\'e de Nantes, IMT Atlantique, IN2P3/CNRS, 4 rue Alfred Kastler, 44307 Nantes cedex 3, France}

\author{Baoyi Chen}
\email{baoyi.chen@tju.edu.cn}
\affiliation{Department of Physics, Tianjin University, Tianjin 300354, China}

\begin{abstract}
We employ the transport model coupled with hydrodynamic equations to study the charmonium dissociation and regeneration in the quark-gluon plasma (QGP) with the large baryon chemical potential in Au-Au collisions at the energies of $\sqrt{s_{NN}}=$ ($39$, $14.5$, $7.7$) GeV. The baryon chemical potential $\mu_B$ is encoded in both Debye mass characterizing the heavy-quark potential and also the equation of state (EoS) of the bulk medium respectively. After considering $\mu_B$-corrections in both heavy quarkonium and the QGP medium, we calculate the nuclear modification factor $R_{AA}$ of charmonium. And find the $\mu_B$ influence on charmonium production at $\sqrt{s_{NN}}$ = 39 and 14.5 GeV is negligible, while the $R_{AA}$ of charmonium is reduced considering $\mu_B$ influence at $\sqrt{s_{NN}}=7.7$ GeV Au-Au collisions. It is crucial for studying charmonium production in low energy and also fixed-target heavy-ion collisions. 
\end{abstract}
\date{\today}

 \maketitle
 
\section{Introduction}

In relativistic heavy-ion collisions, a hot deconfined medium consisting of quarks and gluons, called ``Quark-Gluon Plasma'' (QGP) is believed to be created. 
Heavy quarkonia which are produced in the initial 
parton hard scatterings have been regarded as clean probes of the QGP production~\cite{Matsui:1986dk}. Heavy quarkonium is the bound state of the heavy quark and its antiquark forced by an attractive potential.
In the hot medium, this attractive potential is screened by thermal partons, which results in the 
dissociation of heavy quarkonium and suppression of its production in relativistic heavy-ion collisions~\cite{Satz:2005hx}. The nuclear modification factor $R_{AA}$ is an observable proposed to characterize such suppression.
And various theoretical models have been developed to study the heavy quarkonium evolution and suppression in the hot medium, such as the statistic hadronization model~\cite{Andronic:2003zv}, coalescence hadronization model~\cite{Zhao:2017yan,Chen:2021akx}, transport model~\cite{Yan:2006ve,Zhou:2014kka,Chen:2018kfo,Zhao:2010nk,Du:2015wha,Yao:2020eqy}, open quantum system~\cite{Blaizot:2015hya,Brambilla:2020qwo,Delorme:2022hoo,Miura:2022arv}, time-dependent Schr\"odinger equation~\cite{Katz:2015qja,Islam:2020gdv,Wen:2022utn}, and newly extended Remler equation~\cite{Villar:2022sbv}. The heavy quarkonia not only can be used to probe the QGP properties but also be widely investigated to probe the early state tilted energy deposition and fluctuation of heavy-ion collisions~\cite{Chen:2019qzx,Zhao:2021voa}, and strong electromagnetic and vorticity fields created in non-central heavy-ion collisions~\cite{Alford:2013jva,Marasinghe:2011bt,Chen:2020xsr,Guo:2015nsa,Hu:2022ofv}.

At low energy collisions such as the Beam Energy Scan (BES) program at RHIC~\cite{Bzdak:2019pkr}, NA60+ at SPS~\cite{Dahms:2673280}, and Compressed Baryonic Matter (CBM) at FAIR~\cite{CBM:2016kpk}, the temperatures of the created medium are much lower than that in heavy-ion collisions at top RHIC and LHC energies. While the baryon chemical potential $\mu_B$ in the central regions of low energy collisions enhanced dramatically. It can be much larger than the temperature of the QGP medium, such as the $\mu_B\sim420$ MeV at Au-Au collisions with $\sqrt{s_{NN}}$ = 7 GeV. How such large baryon chemical potential affects heavy quarkonia evolution and production is an interesting and worth-studying question. 
The previous studies based on Hard Thermal and Dense Loop (HTL/HDL) theory show the Debye mass and heavy quark potential are changed in the baryon-rich medium~\cite{Lafferty:2019jpr,Huang:2021ysc,Schneider:2003uz}. 
Recently, based on the time-dependent Schr\"odinger equation, the charmonium dissociation is studied with baryon chemical potential corrected potential~\cite{Tong:2022wpu}. In this paper, we will establish a more realistic model to study carefully the quarkonium dissociation and production in a high baryon density QGP medium which is created in Au-Au collisions at the energies of $\sqrt{s_{NN}}=$ ($39$, $14.5$, $7.7$) GeV.

We first employ the $\mu_B$- and $T$-dependent heavy quark potential to calculate the binding energies and averaged radius of charmonium via the two-body Schr\"odinger equation in Section II. The binding energies are used to estimate the charmonium dissociation in high $\mu_B$-QGP. 
The evolution of the QGP is described by the hydrodynamic model. $\mu_B$ contribution is encoded in the equation of state of hot medium. While the evolution of charmonium is controlled by the Boltzmann equation. Including $\mu_B$ contributions in both heavy quarkonium and the bulk medium, we present the charmonium yield and suppression in section III. A summary is given in section IV.

\section{Theoretical Framework}
\subsection{Heavy quark properties with high baryon density} 
In a vacuum, the mass spectrum of heavy quarkonium is well described with the Cornell potential, $V(r)=-\alpha/r+\sigma r$. At finite temperatures, heavy quark potential is screened. The hot medium effects can be absorbed in the heavy quark potential via the Debye mass $m_D$~\cite{Brambilla:2008cx}. The in-medium potential can be parametrized as~\cite{Lafferty:2019jpr}, 
\begin{eqnarray}
V(r,T) &=& -\alpha\left[m_D+\frac{e^{-m_Dr}}{ r}\right]\nonumber \\
&+&\frac{\sigma}{m_D}\left[2-(2+m_Dr)e^{-m_Dr}\right],
\label{eq-hqp}
\end{eqnarray}
where the parameter {$\alpha=0.4105$ and the string strength $\sigma=0.2\ \rm{(GeV)^2}$} are fixed with the charmonium masses in vacuum~\cite{Hu:2022ofv}. 
At zero baryon chemical potential, the Debye mass can be extracted by fitting the lattice QCD data~\cite{Lafferty:2019jpr}, 
\begin{eqnarray}
m_D(T)&=&g(\Lambda)T\sqrt{{N_c\over 3}+{N_f\over 6}}\nonumber \\
&+&{N_c T g^2(\Lambda)\over 4\pi} \log \left({1\over g(\Lambda)}\sqrt{{N_c\over 3}+{N_f\over 6}} \right) \nonumber\\
&+& \kappa_1Tg^2(\Lambda) + \kappa_2Tg^3(\Lambda)+ \kappa_3Tg^5(\Lambda),
\label{eq-mdt}
\end{eqnarray}
where {$g(\Lambda)$ is the coupling constant depending on the renormalization scale, which can be chosen as $\Lambda=2\pi T$. In this paper, we utilize the four-loop result given in~\cite{Vermaseren:1997fq} with $\Lambda_{QCD}$=0.2 GeV}. The factors of color and flavor are taken as $N_c=N_f=3$. At higher orders, the parameters $\kappa_1$=0.6, $\kappa_1$=-0.23, and $\kappa_3$=-0.007.

\begin{figure}[!htb]
\includegraphics[width=0.35\textwidth]{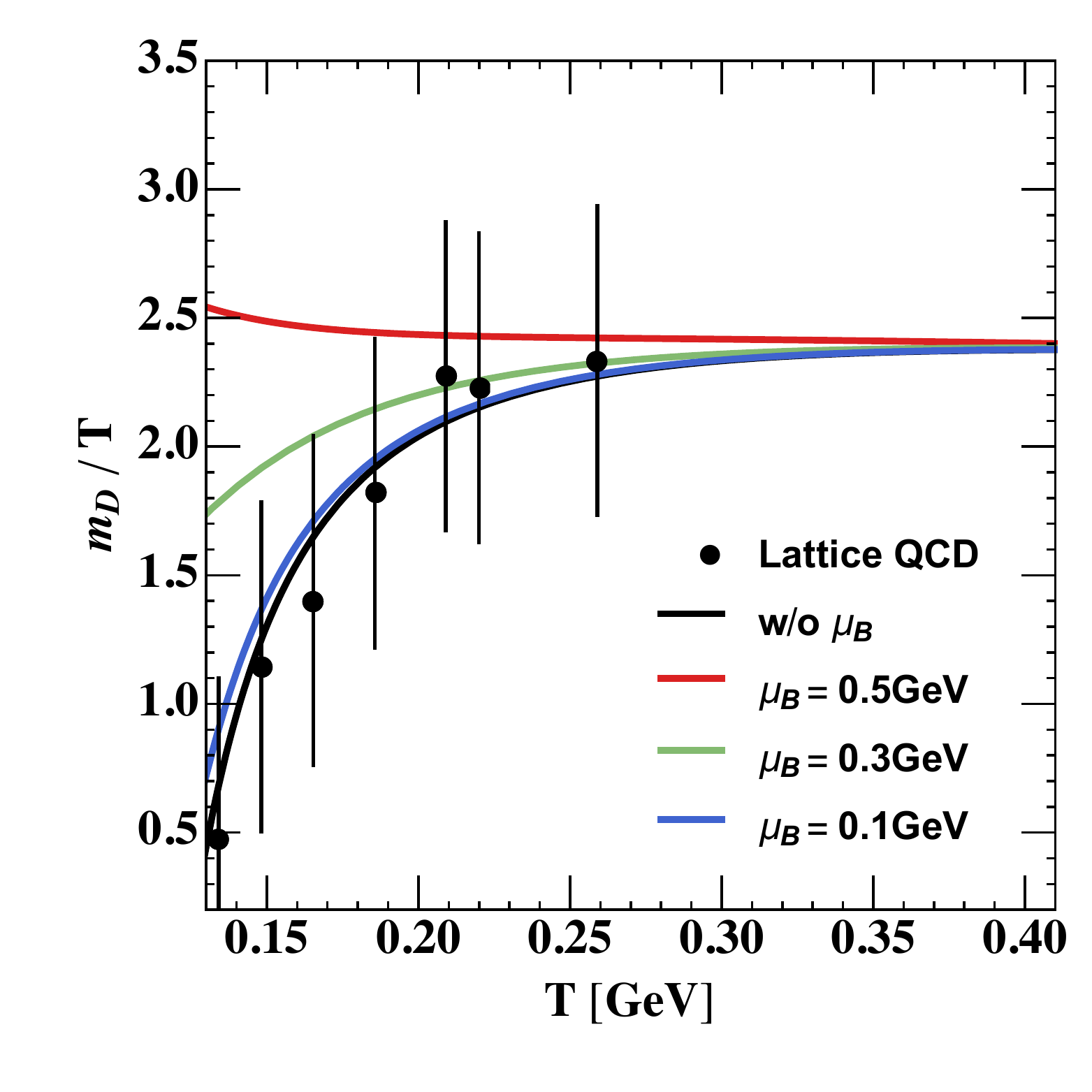}
\caption{The temperature-scaled Debye mass as a function of temperature. Different values of $\mu_B$ are taken to compare with the data which is from lattice QCD results and without the $\mu_B$~\cite{Lafferty:2019jpr}.}
\label{lab-fig1}
\end{figure}

In the high baryon density and hot medium ($T, \mu_B \gg m_q$ with $m_q$ is light quark mass), the leading order HTL/HDL calculations give the Debye screening mass~\cite{Huang:2021ysc,Schneider:2003uz},
\begin{eqnarray}
m_D^2(T,\mu_B)=g^2T^2\left({N_c\over 3}+{N_f\over 6}\right)+g^2\sum_f {\mu_B^2\over 18\pi^2},
\end{eqnarray}
where $\mu_B$ is baryon chemical potential. Due to the lack of lattice data on heavy quark potential at high baryon density regions, we following the study~\cite{Lafferty:2019jpr} to consider the baryon chemical potential through 
\begin{eqnarray}
m_D^2(T,\mu_B)=m_D^2(T)+g^2N_f{\mu_B^2 \over 18\pi^2},
\end{eqnarray}
where $m_D(T)$ is given by lattice results as shown in Eq.(\ref{eq-mdt}). Besides, the renormalization scale in the coupling constant is also modified to $\Lambda=2\pi \sqrt{T^2+\mu_B^2/\pi^2}$. The temperature-scaled Debye mass is plotted in Fig.~\ref{lab-fig1} by taking different values of $\mu_B$. {We can see the Debye mass changes a lot considering the $\mu_B$, especially at low temperature regions.} Taking $m_D(T,\mu_B)$ into Eq.(\ref{eq-hqp}), we obtain the baryon chemical potential related to heavy quark potential.

\begin{figure*}[!tbp]
\centering
\includegraphics[width=0.28\textwidth]{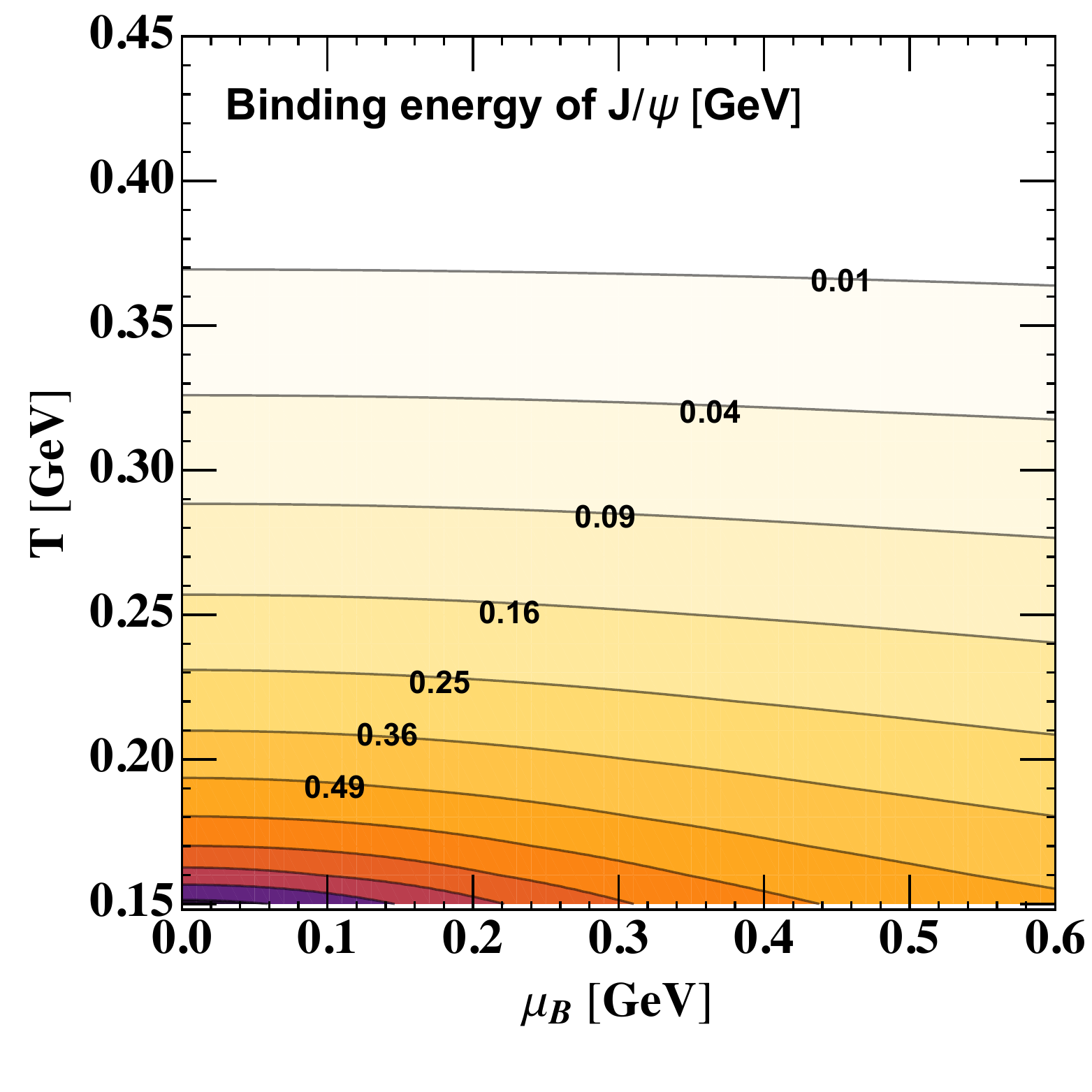}
\includegraphics[width=0.28\textwidth]{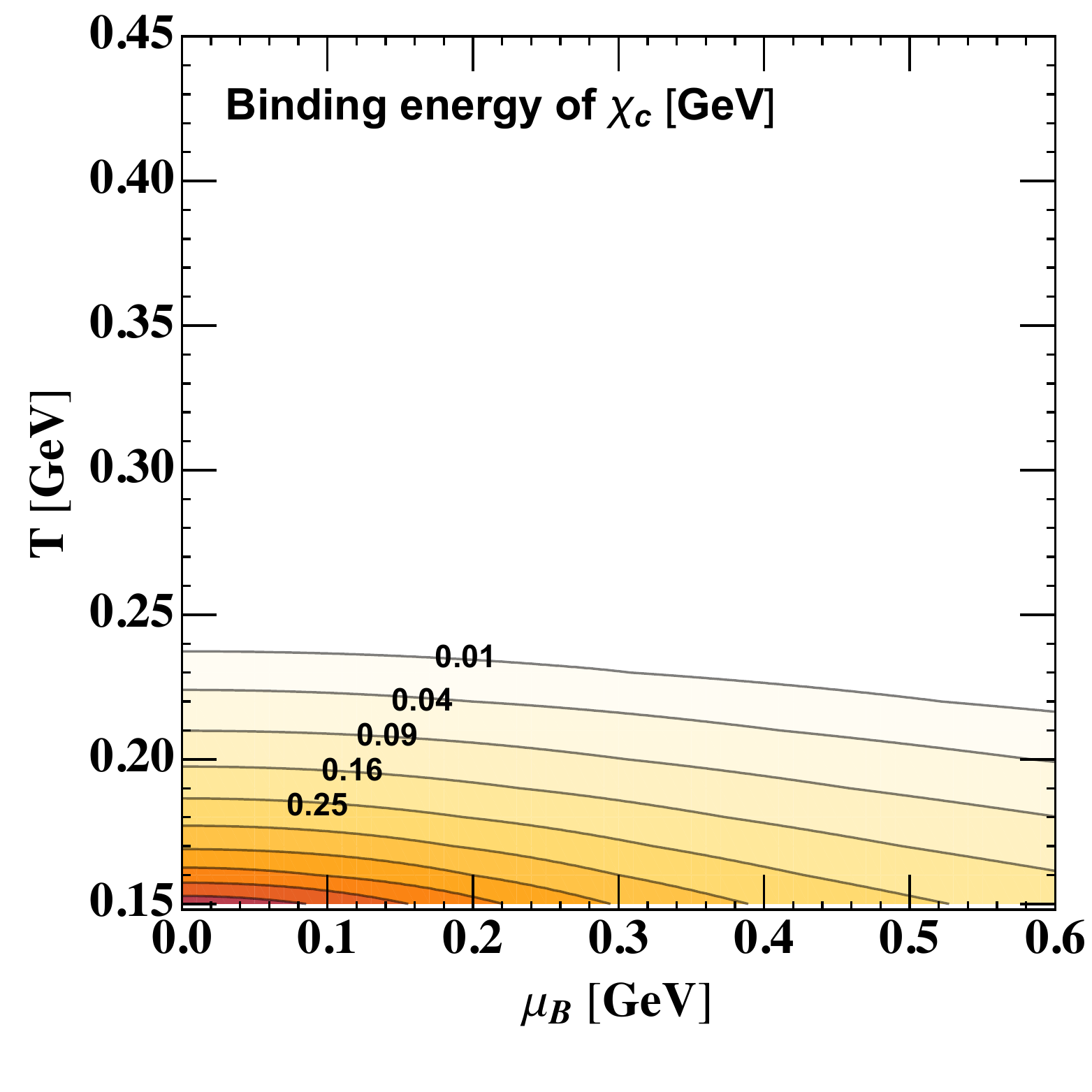}
\includegraphics[width=0.28\textwidth]{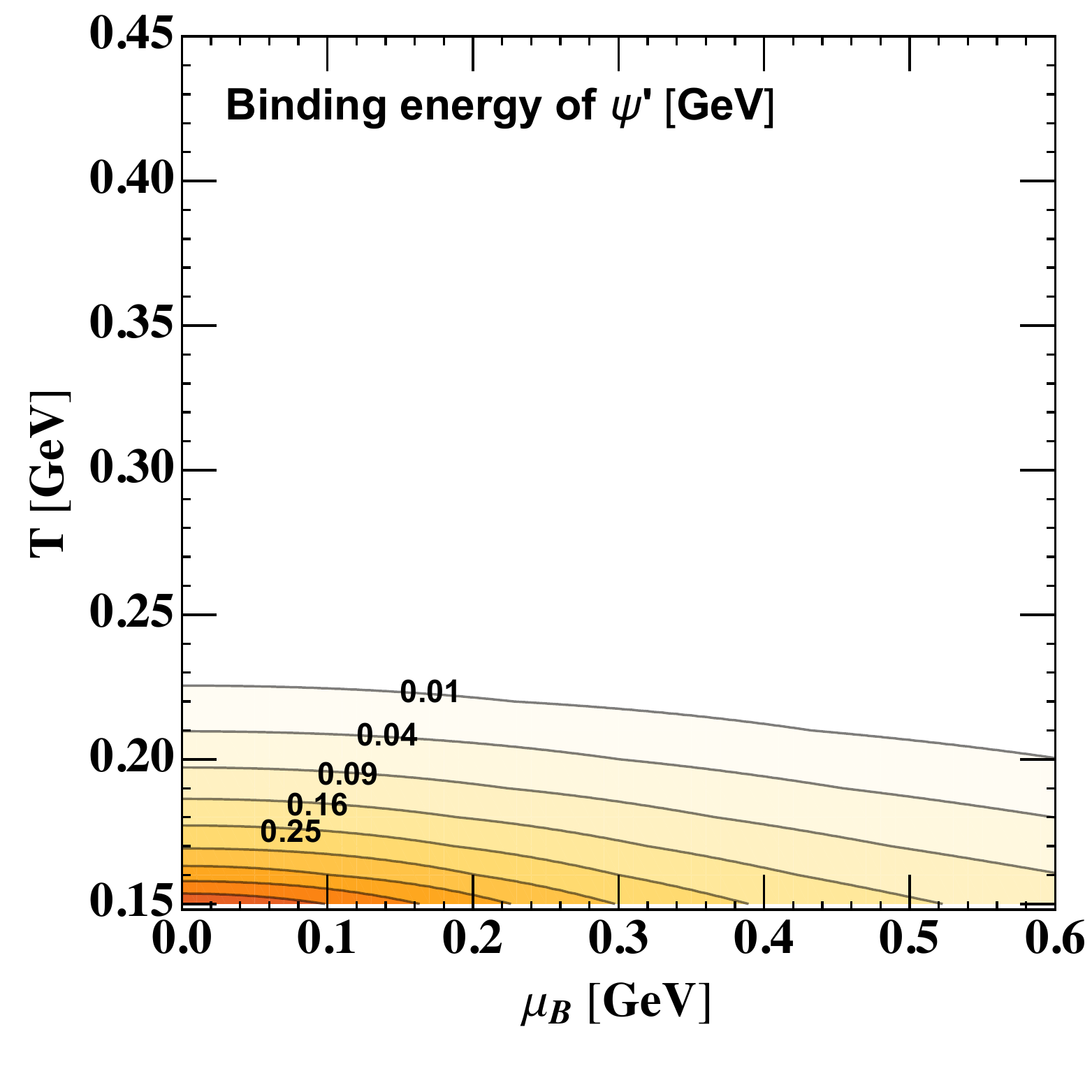}
\caption{Binding energies of charmonium
$(J/\psi, \chi_c, \psi^\prime)$ at different temperatures $T$ and baryon chemical potentials $\mu_B$. 
}
\label{lab-fig2}
\end{figure*}

With ($T$,$\mu_B$)-dependent heavy quark potential, one can calculate the in-medium binding energies of charmonium. As the charm quark mass is relatively large compared with the inner motion of charm quarks in the bound state, we neglect the relativistic effect and employ the Schr\"odinger equation to calculate their binding energies via $\epsilon_\psi(T,\mu_B)=E-V(r=\infty,T,\mu_B)$. The potential Eq.(\ref{eq-hqp}) is a central potential, the radial part of the Schr\"odinger equation is separated to be, 
\begin{eqnarray}
&&\left[{1\over 2\mu} \left(-{d^2\over d^2r}-{2\over r}{d\over dr}+{l(l+1)\over r^2} \right)+V(r,T,\mu_B) \right]R_{nl}(r)\nonumber\\
&&=ER_{nl}(r),
\end{eqnarray}
where $R_{nl}(r)$ is the radial wave function labeled with the radial and angular quantum number $(n,l)$. 
$r=|{\bf r}_2-{\bf r}_1|$ is the distance between charm and anti-charm quarks located at the positions of ${\bf r}_1$ and ${\bf r}_2$. The reduced mass is $\mu=m_c/2$ with charm mass $m_c=1.5$ GeV. Here we consider three charmonium states $(J/\psi, \chi_c, \psi^\prime)$. Their binding energies at different $\mu_B$ and $T$ are plotted in Fig.~\ref{lab-fig2}. As one can see, the binding energies of charmonium are reduced considering the baryon chemical potential $\mu_B$. 

In the meantime, charmonia will suffer dynamic dissociation in the QGP, such as gluo-dissociation and inelastic scattering with thermal partons. For the deeply bounded charmonium state, the gluo-dissociation plays an important role, which cross section can be obtained by the operator-production-expansion (OPE) method~\cite{Peskin:1979va,Bhanot:1979vb}, 
\begin{eqnarray}
\sigma_{1s}(\omega)&=&A_0 {(x-1)^{3/2}\over x^5},\nonumber\\
\sigma_{\chi_c}(\omega)&=&4A_0 {(x-1)^{1p}(9x^2-20x+12)\over x^7},\nonumber\\
\sigma_{2s}(\omega)&=&16A_0 {(x-1)^{3/2}(x-3)^2\over x^7},
\end{eqnarray}
where $\omega$ is the gluon energy.   $x\equiv\omega/\epsilon_\psi$ is the ratio of gluon energy and the binding energy $\epsilon_\psi$ of charmonium states. The coefficient $A_0=2^{11}\pi/(27\sqrt{m_c^3\epsilon_\psi})$ is a constant factor. At finite temperatures and baryon chemical potentials, heavy quark potential becomes weaker for the charmonium states with larger geometry sizes.  We use the geometry scale to obtain the in-medium inelastic cross sections, \begin{eqnarray}
\sigma(T,\mu_B)=\sigma(T=0, \mu_B=0)\frac{\langle r^2\rangle_{T,\mu_B}}{ \langle r^2\rangle_{T=0,\mu_B=0}},
\end{eqnarray}
where the mean-square-radius $\langle r^2\rangle$ of charmonium in a vacuum and the finite $(T,\mu_B)$ can be calculated with the Schr\"odinger equation by taking the vacuum Cornell potential and the screened potential $V(r, T,\mu_B)$, respectively.

\subsection{QGP medium with high baryon density}
\begin{figure}[!htb]
\centering
\includegraphics[width=0.35\textwidth]{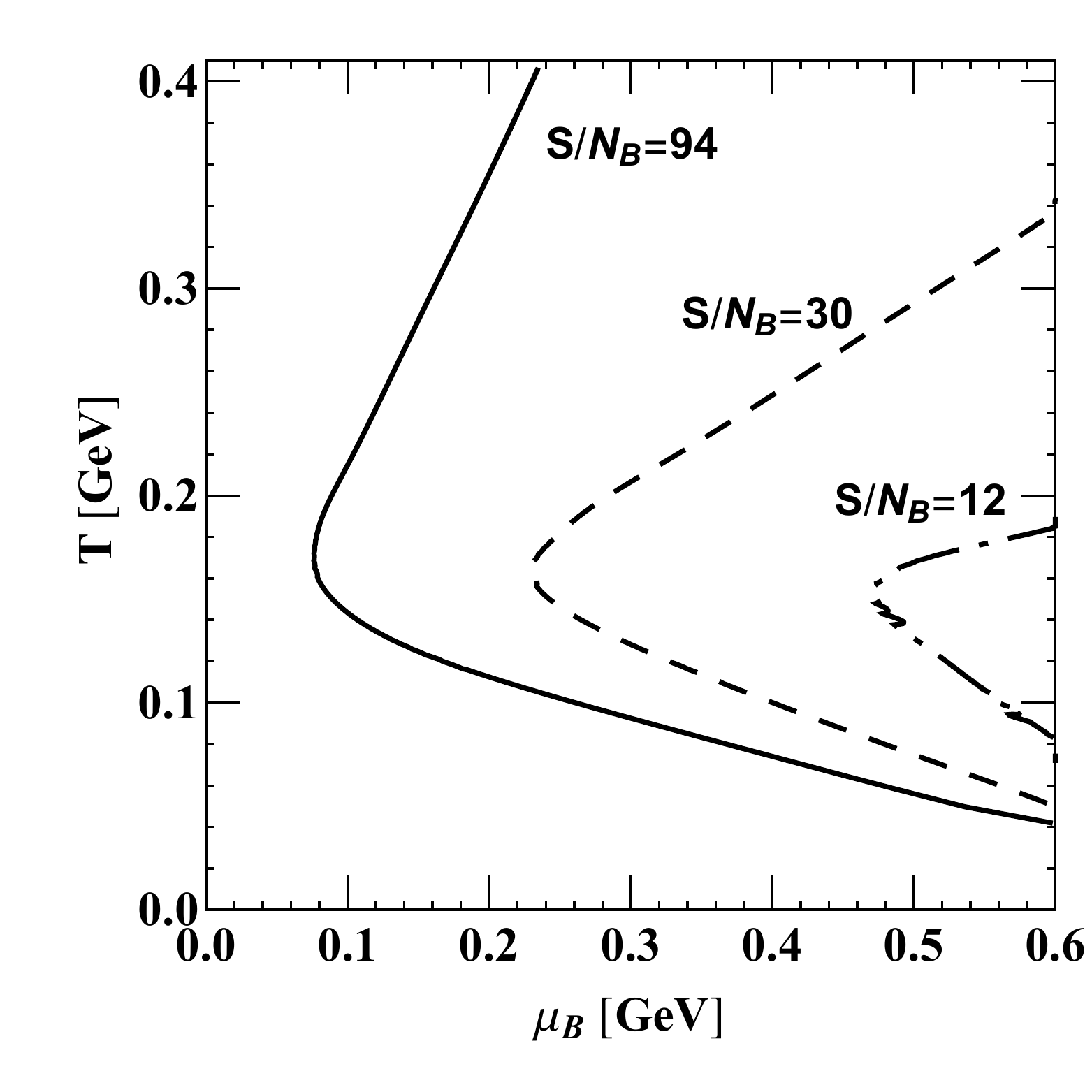}
\caption{ {The relations between temperature and baryon chemical potential at the fixed values of the entropy $S$ over baryon number $N_B$. Three lines labelled with $S/N_B=(94,30,12)$ correspond to the medium produced at Au-Au collisions with $\sqrt{s_{NN}}=39,14.5, 7.7$ GeV, respectively~\cite{Guenther:2017hnx}.  }
}
\label{lab-fig3}
\end{figure}

\begin{figure}[!htb]
\centering
\includegraphics[width=0.35\textwidth]{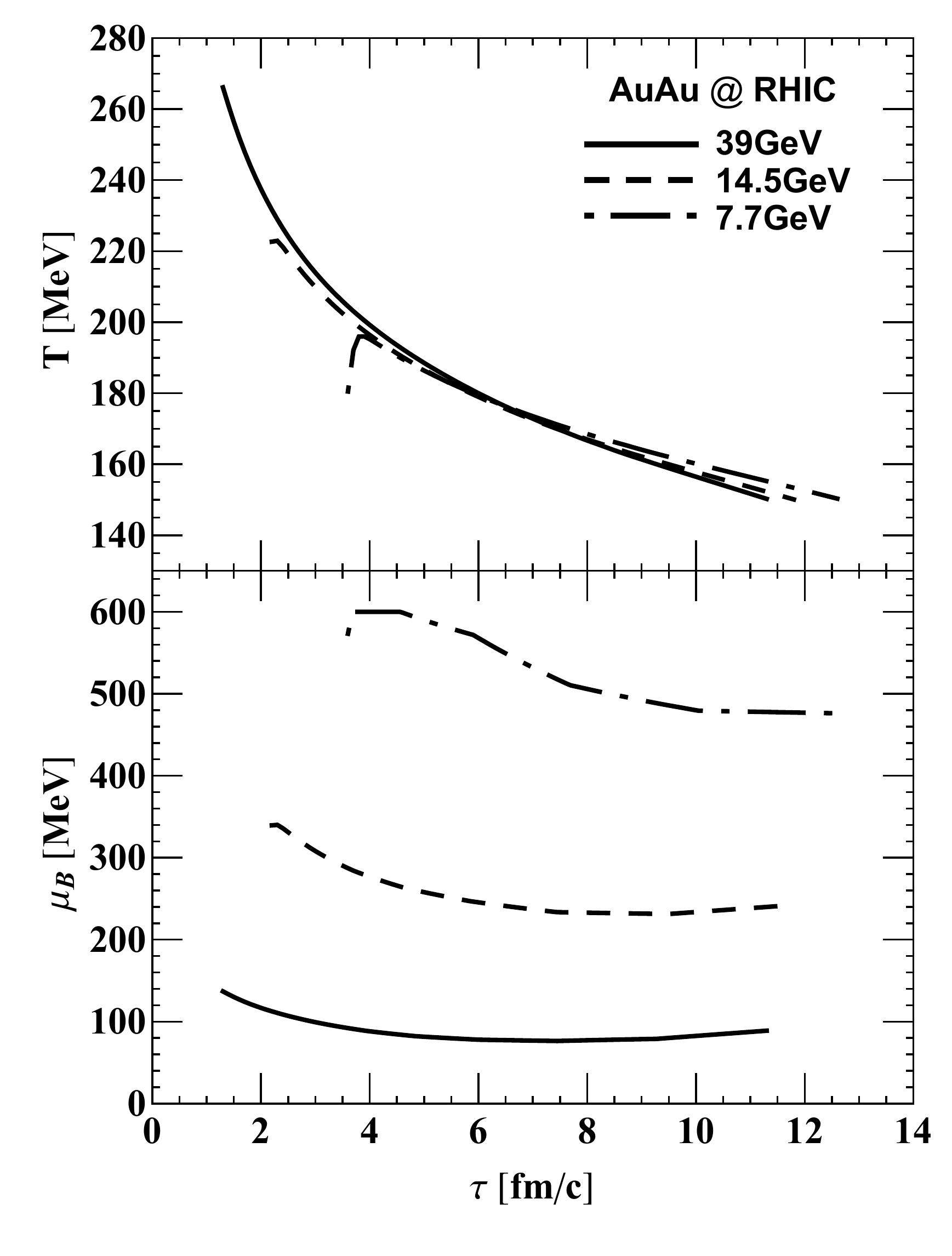}
\caption{ (Upper panel) Time evolution of the temperature at the center of the medium in the most central Au-Au collisions at $\sqrt{s_{NN}}$= 39, 14.5, and 7.7 GeV.
(Lower panel) Time evolution of the baryon chemical potential at the center of the medium at three collision energies.
}
\label{lab-fig4}
\end{figure}

The quark matter created in heavy-ion collisions is nearly a perfect fluid with a small ratio of the viscosity over entropy density $\eta/s$ ~\cite{Teaney:2003kp,Lacey:2006bc}. The space-time evolution of the medium can be described such as MUSIC model, where hydrodynamic equations and the conservation equation for net baryon density are~\cite{Schenke:2010nt},
 \begin{eqnarray}
 \label{hydro}
\partial_\mu T^{\mu \nu} &=& 0, \nonumber\\
\partial_\mu J_B^\mu &=& 0.
\end{eqnarray}
The energy-momentum tensor $T^{\mu \nu}$ and the net baryon current $J_B^\mu$ are expressed as
 \begin{eqnarray}
 \label{tuvb}
T^{\mu \nu} &=& \epsilon u^\mu u^v-(P+\Pi)\Delta^{\mu \nu}+\pi^{\mu \nu}, \nonumber\\
J_B^\mu &=& n_Bu^\mu + q^\mu,
\end{eqnarray}
where $\epsilon$ and $P$ are the medium energy density and the pressure respectively. $\Pi$ is the bulk viscous pressure, $\pi^{\mu \nu}$ is the shear stress tensor. $\Delta^{\mu \nu}= g^{\mu \nu}-u^\mu u^\nu$ is the projection tensor. The net baryon density and the baryon diffusion current are written as $n_B$ and $q^\mu$. In the following calculations, 
we take the ratio of shear viscosity to entropy density as a constant $\eta/s=0.08$ and neglect the bulk viscosity. The equation of state at finite $\mu_B$ and $T$ are taken as NEOS-B~\cite{Monnai:2019hkn,Monnai:2021kgu,Guenther:2017hnx}. The initial time $\tau_0$ of the hot medium can be estimated from the overlap time of the two colliding nuclei~\cite{Denicol:2018wdp,Wu:2021fjf,Shen:2017bsr}. Here we take $\tau_0$= 1.3, 2.2, 3.6 $\rm{fm/c}$ for Au-Au collisions at $\sqrt{s_{NN}}=39,14.5, 7.7$ GeV respectively. The $T-\mu_B$ relations with different values of entropy over baryon number $S/N_B$ are plotted in Fig.~\ref{lab-fig3}. With the final charge multiplicities from experiments, we can determine the initial temperatures of the hot medium at different collision energies. The time evolution of $T$ and $\mu_B$ at the center of the medium are plotted in Fig.~\ref{lab-fig4}. We can see the temperature and $\mu_B$ decrease as the medium expands.

\subsection{Charmonium transport equation}
Charmonium dynamical evolution in the hot medium can be 
described with the Boltzmann transport equation. The 
dissociation process and the recombination from $c$ and $\bar c$ quarks in the QGP are encoded in the decay rate $\alpha$ and regeneration rage $\beta$ of charmonium bound states in the below equation. The phase space distribution $f_\psi$ of charmonium satisfy the equation~\cite{Yan:2006ve,Zhou:2014kka,Chen:2018kfo}, 
\begin{align}
&\left[ \cosh(y-\eta)\frac{\partial}{\partial \tau} +\frac{\sinh(y-\eta)}{\tau}\frac{\partial}{\partial \eta}+{\bf v}_T\cdot \nabla_T \right] f_{\psi} \nonumber \\&=-\alpha f_{\psi}+\beta,
\end{align}
where $\eta=1/2\ln[(t+z)/(t-z)]$ and $y=1/2\ln[(E+p_z)/( E-p_z)]$ are the spatial and momentum rapidity, respectively. ${\bf v}_T={\bf p}_T/E_T$ is the transverse velocity of charmonium and $E_T=\sqrt{m_{\psi}^2+{\bf p}_T^2}$ is the transverse energy. 
On the right-hand side, $\alpha$ and $\beta$ are the loss and gain terms. The decay rate $\alpha$ depends on the charmonium dissociation cross sections and the parton densities. While the regeneration rate depends on the charm densities and the recombination cross section~\cite{Zhao:2022ggw},
\begin{align}
\label{alphabeta}
\alpha({\bm p},x) &= {1\over 2E_\psi}\int{d^3{\bm p}_g\over(2\pi)^32E_g}W_{g\psi}^{c\bar c}(s)f_g({\bm p}_g,x) \Theta(T(x)-T_c),\nonumber\\
\beta({\bm p},x) &= {1\over 2E_\psi}\int {d^3{\bm p}_g\over(2\pi)^32E_g}{d^3{\bm p}_c \over(2\pi)^32E_c}{d^3{\bm p}_{\bar c}\over(2\pi)^32E_{\bar c}} F_c F_r \nonumber \\
&\quad\times W_{c\bar c}^{g\psi}(s) f_c({\bm p}_c,x)f_{\bar c}({\bm p}_{\bar c},x)\nonumber\\
&\quad \times \Theta(T(x)-T_c)(2\pi)^4\delta(p+p_g-p_c-p_{\bar c}),
\end{align}
where $E_g$ and $E_\psi$ are the energies of gluon and charmonium respectively. ${\bf p}$ and 
${\bf p}_g$ are the momentum of charmonium and the gluon, respectively. Gluon density $f_g$ satisfies Bose distribution. 
$W_{g\psi}^{c\bar c}(s)$ is the dissociation probability constructed by the cross section $\sigma_{g\psi}^{c\bar c}(s)$ and the gluon flux. It is connected with the recombination 
probability $W_{c\bar c}^{g\psi}$ via the 
detailed balance. 
The step function $\Theta(T(x)-T_c)$ ensures the dissociation process happens in the QGP with $T>T_c$. $T_c$ is the temperature of the deconfinement phase transition and depends on $\mu_B$~\cite{Randrup:2006nr}. With the finite baryon chemical potentials in Au-Au collisions at $\sqrt{s_{NN}}=(39, 14.5, 7.7)$ GeV, the value of $T_c$ is determined to be around (0.165, 0.156, 0.140) GeV respectively in the EoS ~\cite{Randrup:2006nr}. $f_c$ and $f_{\bar c}$ are the Fermi-distributions of charm and anti-charm quarks with the conservation of the total number in the medium. 
At low collision energies, charm quarks can not reach kinetic thermalization immediately in the medium. The non-thermalization effect is included with a 
relaxation factor $F_r=1-e^{-\tau/\tau_r}$, where $\tau_r\simeq 7\  fm/c$ characterizes the time scale of medium local thermalization~\cite{Grandchamp:2003uw,Zhao:2007hh}. Due to a small number of charm pairs produced in the medium, charm quark distribution is described with the canonical ensemble. This effect can be included in the regeneration rate via the canonical factor 
$F_c=1+{1/(dN_{c\bar c}/dy)}$\cite{Liu:2013via,Gorenstein:2000ck,Zhao:2022ggw} where $dN_{c\bar c}/dy$ is the charm pairs in a unit rapidity. 

\begin{figure}[!htb]
\includegraphics[width=0.35\textwidth]{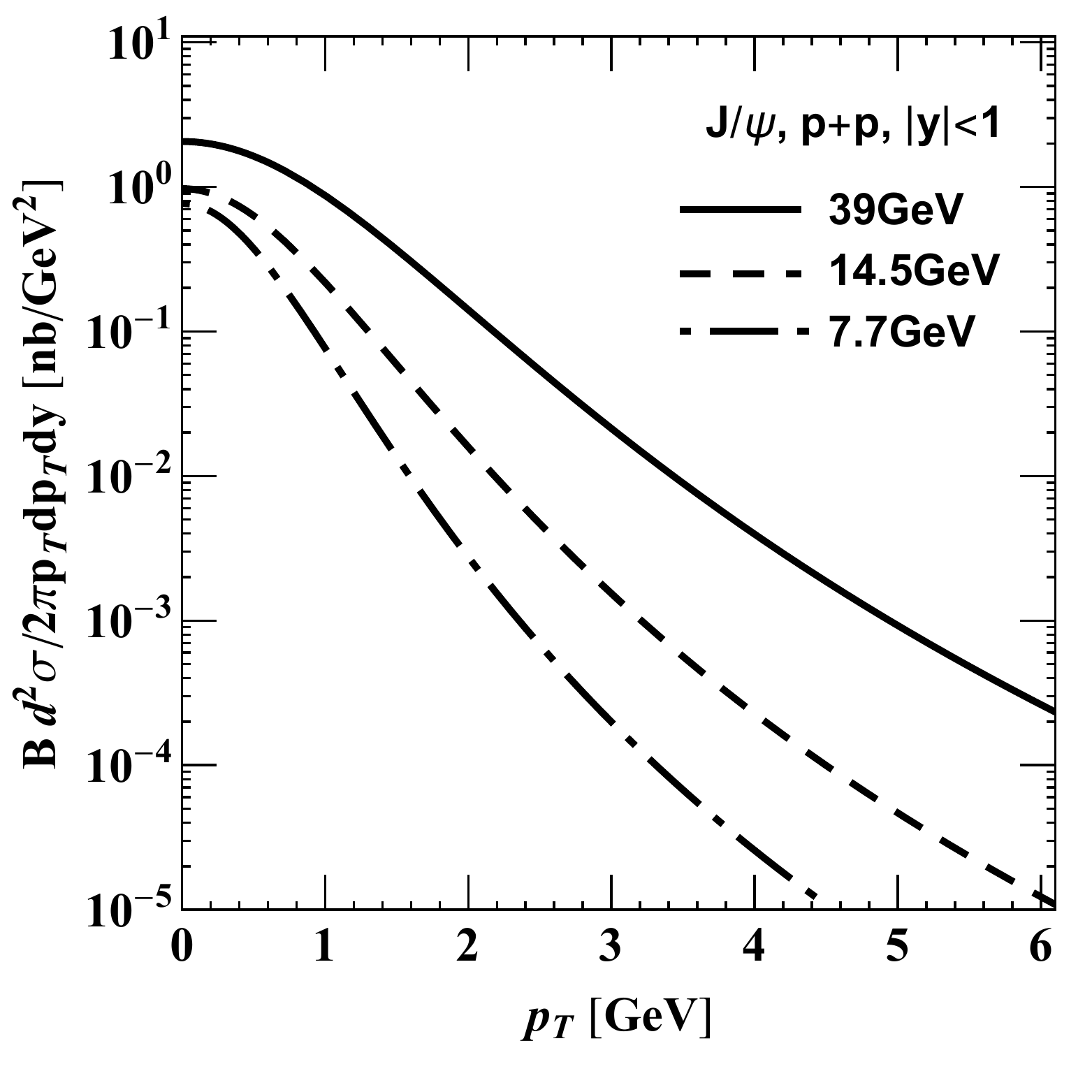}
\caption{ The $p_T$ spectrum of $J/\psi$ produced in the mid-rapidity p-p collisions at $\sqrt{s_{NN}}$ = 39, 14.5, and 7.7 GeV. The value of the branching ratio is $B(J/\psi\to e^+e^-)=5.97\pm0.03\%$~\cite{ParticleDataGroup:2020ssz}. 
}
\label{lab-pp-pt}
\end{figure}

The initial momentum distribution of 
charmonium in nuclear collisions is extracted from the measurements in proton-proton (p-p) collisions. Charmonium distribution in p-p collision is fitted with the formula~\cite{Zha:2015eca}, 
\begin{align}
\label{initial}
{d^2\sigma^{J/\psi}_{pp}\over 2\pi p_Tdp_Tdy}= {a\over 2\pi\langle p_T^2\rangle}\left(1+b^2{p_T^2\over \langle p_T^2\rangle}\right)^{-n} {d\sigma^{J/\psi}_{pp}\over dy}
\end{align}
where the parameters are $a=2b^2(n-1),\ b=\Gamma(3/2)\Gamma(n-3/2)/\Gamma(n-1)$ and $n=3.93\pm0.03$. {At $\sqrt{s_{NN}}$= 39, 14.5, and 7.7 GeV, the mean transverse momentum square of charmonium is taken as $\langle p_T^2\rangle$= 1.46, 0.77, and 0.45 $\rm{(GeV)^2}$. The differential cross section of $J/\psi$ is taken to be $d\sigma^{J/\psi}_{pp}/dy$= 150.7, 37.6, and 17.2 nb in central rapidity~\cite{Zhao:2022ggw,Zha:2015eca}.} The momentum distribution of $J/\psi$ at three collision energies are plotted in Fig.~\ref{lab-pp-pt}. We take the same initial distribution for the excited states. 
The initial spatial distribution of charmonium in Au-Au collisions is proportional to the product of two thickness function $dN_\psi/d{\bf x}_T\propto T_A({\bf x}_T-{\bf b}/2)T_B({\bf x}_T+{\bf b}/2)$ where ${\bf b}$ and ${\bf x}_T$ are the impact parameter and the transverse coordinate respectively. 

\begin{figure}[!htb]
\centering
\includegraphics[width=0.35\textwidth]{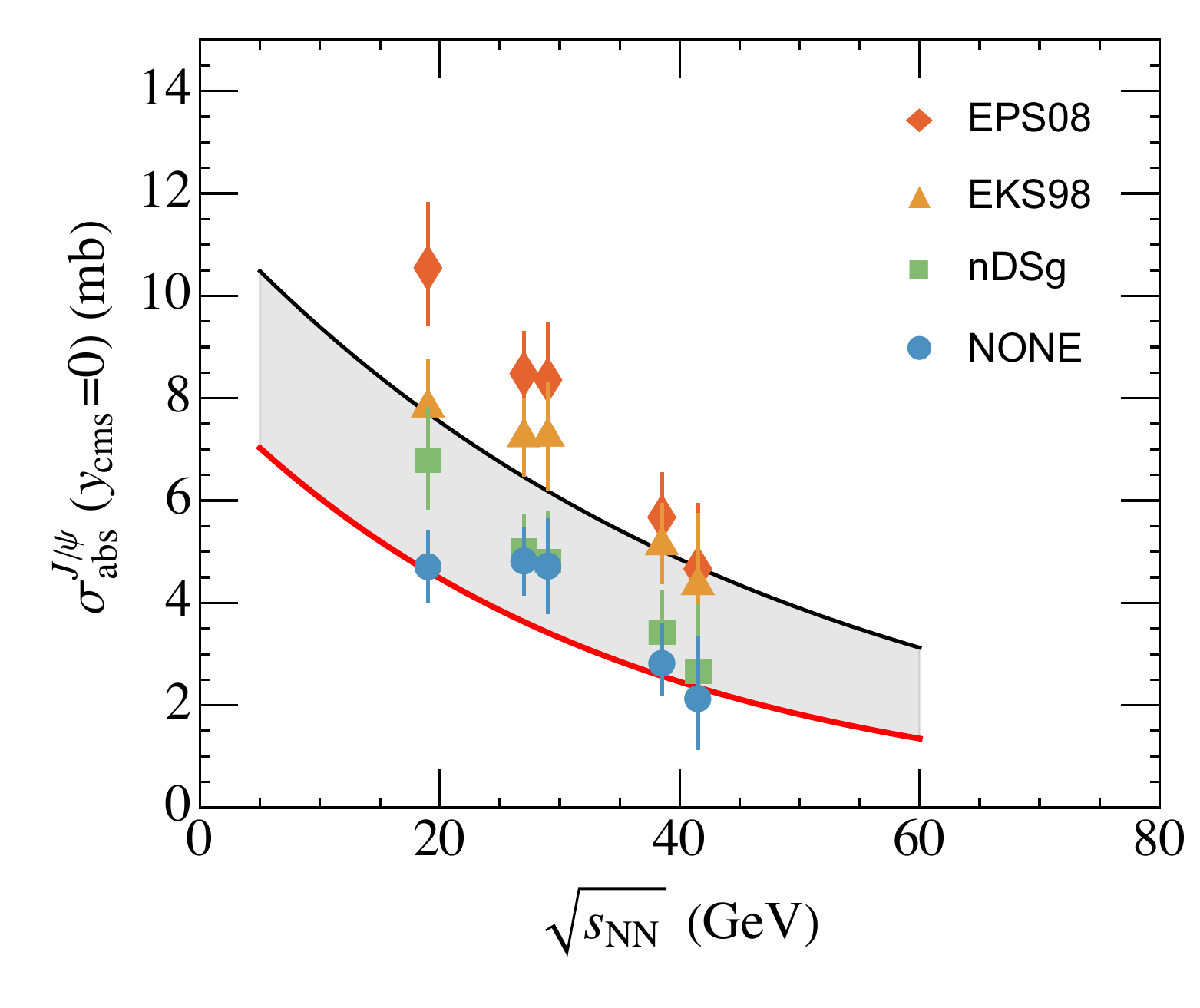}
\caption{ Energy dependence of the nuclear absorption cross sections of $J/\psi$ at mid-rapidity. 
\sred{Different values of the nuclear absorption cross section are extracted based on different models (EPS08, EKS98, nDSg)~\cite{Lourenco:2008sk}, plotted with dots. The line with ``NONE''  
represents minimal cross sections in the limit without the (anti-)shadowing effect. 
These values of 
$\sigma_{abs}^{J/\psi}$ are fitted with a band.}  }
\label{lab-nabs}
\end{figure}

The cold nuclear effects such as nuclear absorption, Cronin effect, and Shadowing effect are included. 
The nuclear absorption survival probability depends on the absorption cross section, which depends on where the state is formed and how far it travels through nuclear matter~\cite{Vogt:2001ky}. 
At LHC energies, very few nucleons stay in this central region resulting in a negligible absorption effect. 
However, the effective absorption cross section increases with decreasing the colliding energy because the quarkonium spends much time traveling through the nuclear matter~\cite{Vogt:2001ky}. The absorption cross sections for the ground state $J/\psi$ at different colliding energies are fixed with the data from p-A collisions~\cite{Brambilla:2010cs,Lourenco:2008sk,Chatterjee:2022ssu}. 
{To explain the experimental data in p-A collisions which are the joint effect of the shadowing factor and the nuclear absorption cross section. When taking different values of the shadowing factor from other models such as EPS08, EKS98, nDSg, one also needs 
to take different absorption cross sections to fit the same experimental data in p-A collisions, please see the dots in Fig.\ref{lab-nabs}. 
In this study, aiming to consider the uncertainty of $\sigma_{abs}^{J/\psi}$ induced by 
the shadowing factors from different models, we fits the $\sigma_{abs}^{J/\psi}$ (shown as dots in Fig.\ref{lab-nabs}) with a band. 
The upper limit of the band gives $\sigma_{abs}^{J/\psi}$= 5.2, 8.9, and 9.7 mb at $\sqrt{s_{NN}}$ = 39, 14.5, and 7.7 GeV, while the lower limit of the band gives $\sigma_{abs}^{J/\psi}$= 2.5, 5.2, and 6.5 mb at $\sqrt{s_{NN}}$ = 39, 14.5, and 7.7 GeV. 
For the excited states, their absorption cross sections} are obtained from $\sigma_{abs}^{J/\psi}$ via the geometry scale~\cite{Zhao:2022ggw}. For the Cronin effect, the additional transverse momentum from the multi-scattering with surrounding nucleons is considered with an increase in the mean transverse momentum square $\langle p_T^2\rangle+a_{gN}\langle L\rangle$ of $J/\psi$~\cite{Huefner:2002tt}. 
The Cronin parameter $a_{gN}$ is the averaged transverse momentum square obtained from the scattering within a unit of length of nucleons, and $\langle L\rangle$ is the mean trajectory length of the two gluons before fusing into heavy quarkonium. We take $a_{gN}$= 0.08, 0.077, and 0.073 $\rm{GeV^2/fm}$ for Au-Au collisions at $\sqrt{s_{NN}}$ = 39, 14.5, and 7.7 GeV~\cite{Zhao:2022ggw}. 
The shadowing factor is simulated by the EPS09 package~\cite{Helenius:2012wd}.
\begin{figure}[!htb]
\centering
\includegraphics[width=0.35\textwidth]{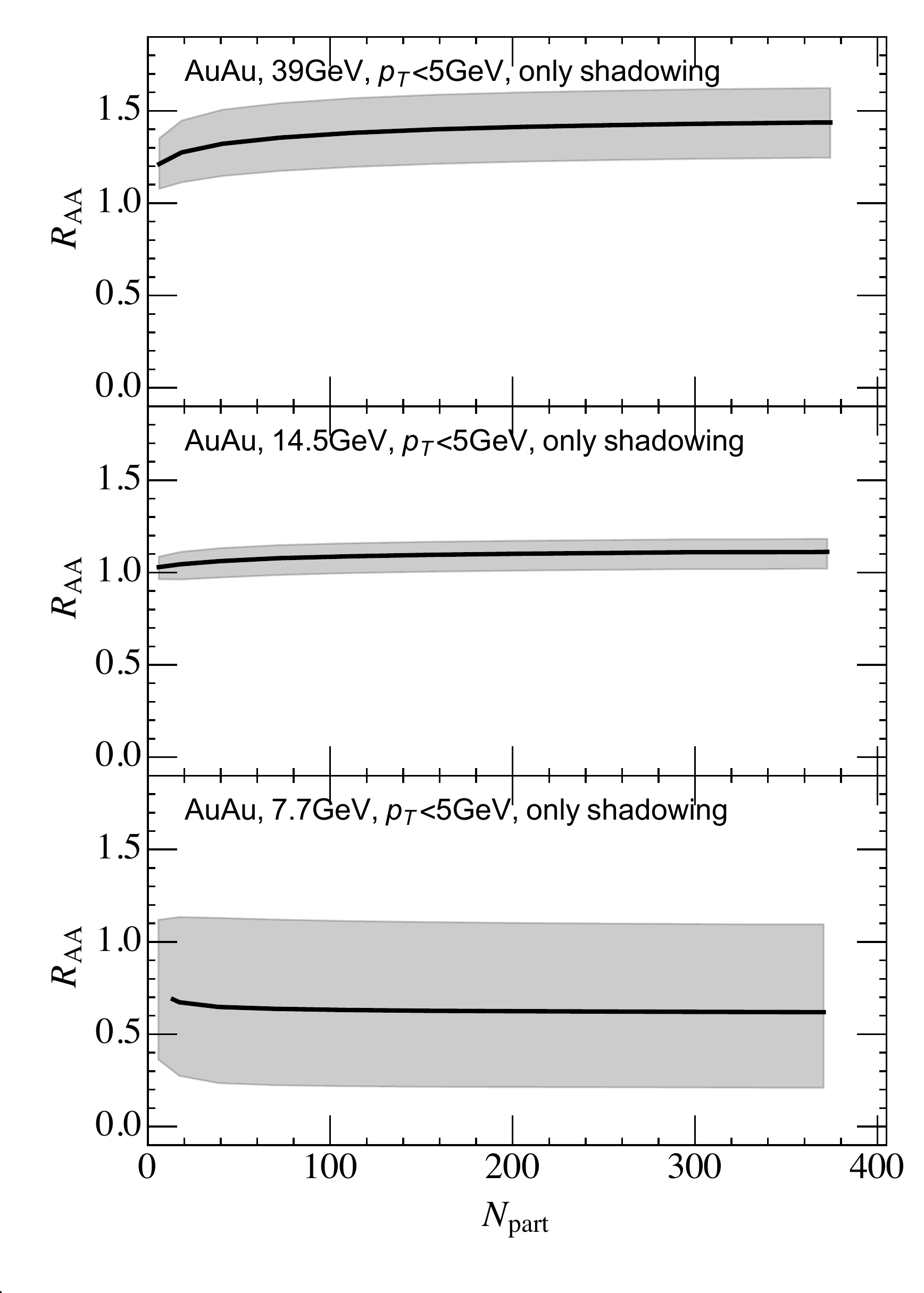}
\caption{$R_{AA}$ of $J/\psi$ with only the shadowing effect. \sred{Different bands in $R_{AA}$ 
are induced by the uncertainty in the shadowing factor taken from other 
models~\cite{Lourenco:2008sk}.} }
\label{lab-shadowing}
\end{figure}

\section{Numerical results}

In this study, we do not include the evolution and dissociation of charmonia in the hadronic phase, which gives a small and additional suppression of charmonium via scattering with hot $\pi$ and $\rho$ mesons~\cite{Lin:1999ad,Haglin:2000ar,Chen:2015ona}. The baryon chemical potential only affects the relative yield of $\pi^+$ and $\pi^-$~\cite{STAR:2017sal}, however, the inelastic scattering between charmonium and $\pi$ meson is charge-independent. So, the difference induced by baryon chemical potential in the hadronic phase can be neglected.
\begin{figure}[!htb]
\centering
\includegraphics[width=0.35\textwidth]{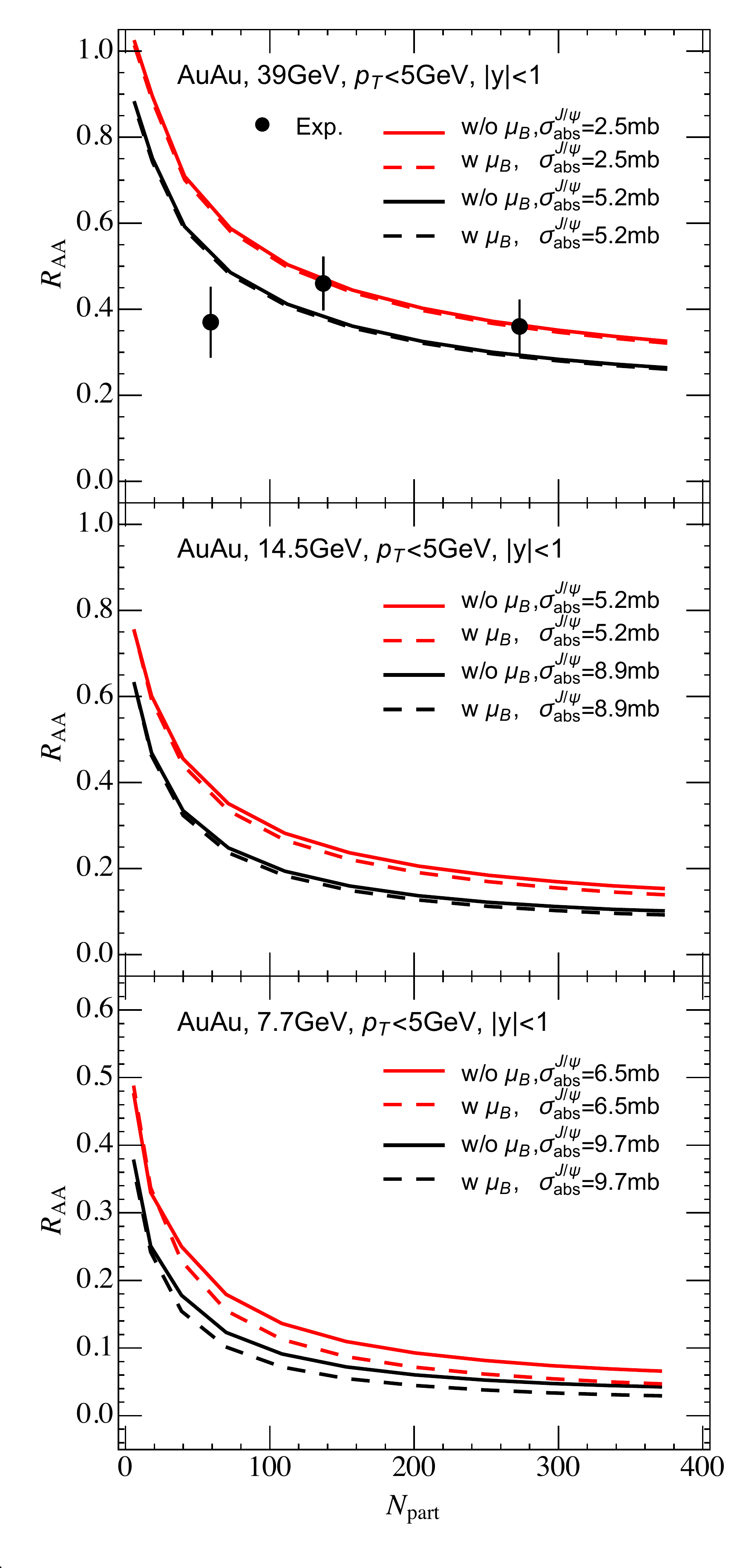}
\caption{ The nuclear modification factor of $J/\psi$ as a function of number of participants $N_{part}$ in $\sqrt{s_{NN}}$=39, 14.5, and 7.7 GeV Au-Au collisions. Solid and dashed lines are results without and with the $\mu_B$ effect. Red and black lines are with different nuclear absorption cross sections. The experimental data is from~\cite{STAR:2016utm}.
}
\label{lab-raa-naprt}
\end{figure}

As we discussed before, the baryon chemical potential affects not only the Debye mass but also the evolution of the QGP medium. With the transport equation for charmonium evolutions and the hydrodynamic model for bulk medium evolutions, we study the charmonium nuclear modification factors at low collision energies.

First, let us see the yield of $J/\psi$ with only the shadowing effect. The results are shown in Fig.~\ref{lab-shadowing}. Due to the (anti-)shadowing effect at 39 and 14.5 GeV, and EMC effect at 7.7 GeV~\cite{Helenius:2012wd}, we can see a clear enhancement of $J/\psi$ at 39 and 14.5 GeV, but a strong suppression at 7.7 GeV. \sred{The very narrow uncertainty band of $R_{AA}$ at 14.5 GeV comes from the well-constrained gluon shadowing factor~\cite{Helenius:2012wd}}. Next, we turn on all cold and hot nuclear matter effects. $J/\psi$ nuclear modification factors at $\sqrt{s_{NN}}$=14.5, 39, and 7.7 GeV Au-Au collisions are also calculated and shown in Fig.~\ref{lab-raa-naprt}. 
At $\sqrt{s_{NN}}=39$ GeV, the regeneration process has been included and contributes around 50\% of the final production in the central collisions. With a nuclear absorption cross sections small $\sigma_{abs}^{J/\psi}$, the suppression is weak no doubt and gives to a large $R_{AA}$.

Although the baryon chemical potential $\mu_B$ can reach 250 and 100 MeV at $\sqrt{s_{NN}}=14.5$ and 39 GeV, \sred{the influence on the charmonium production is very small because of 
the high temperature of the medium. The $\mu_B$ effect changes the heavy quark potential 
via the form $\mu_B/T$. Only at low temperatures and large baryon chemical potential such as 
in $\sqrt{s_{NN}}=7.7$ GeV Au-Au collisions, the Debye mass can be increased by around 50\% at $T_c$ after including the $\mu_B/T$-correction.} After considering both $\mu_B$ corrections in heavy quark potential and the EoS of the medium, $J/\psi$ nuclear modification factor is slightly reduced by around 30\% in most central collisions as shown in Fig.~\ref{lab-raa-naprt}. 
The regeneration process is negligible in this collision energy due to the very small charm density and far from the thermal distribution.

\section{Summary}

We study the charmonium evolutions in the baryon-rich medium produced in the Au-Au collisions at $\sqrt{s_{NN}}$= 39, 14.5, and 7.7 GeV. The $\mu_B$ contribution increases the Debye mass, which in turn reduces heavy quark potential and the in-medium binding energies of the charmonium state. In this study, we consider $\mu_B$ corrections in both heavy quark potential and also the equation of state of the QGP medium. By taking realistic values of the initial baryon chemical potential and the temperature, we calculate the nuclear modification factor of $J/\psi$. It shows the $\mu_B$ influence on charmonium production at $\sqrt{s_{NN}}$ = 39 and 14.5 GeV is negligible, while the $R_{AA}$ of charmonium reduced almost 30\% in the central collisions considering $\mu_B$ influence at $\sqrt{s_{NN}}=7.7$ GeV Au-Au collisions. The suppression of $J/\psi$ caused by high baryon density is even larger than the temperature effect. This is interesting and crucial for studying charmonium production in low-energy and also fixed-target heavy-ion collisions.

\vspace{1cm}
\noindent {\bf Acknowledgement}: This work is supported by the National Natural Science Foundation of China
(NSFC) under Grant Nos. 12175165. This study has, furthermore, received funding from the European Union’s Horizon 2020 research and innovation program under grant agreement No. 824093 (STRONG-2020).


\end{document}